\documentclass[12pt]{article}%
\usepackage{amsmath}
\usepackage{graphicx}
\usepackage{amsfonts}
\usepackage{amssymb}%
\providecommand{\U}[1]{\protect\rule{.1in}{.1in}}
\begin{document}

\title{Inertial spin alignment in a circular magnetic nanotube}
\author{G. Bergmann, R. S. Thompson, and J. G. Lu\\Department of Physics and Astronomy\\University of Southern California\\Los Angeles, California 90089-0484\\e-mail: bergmann@usc.edu}
\date{2015-03-03}
\maketitle

\begin{abstract}
In Co-nanotubes with a curling magnetization, the orbital motion of the
conduction electrons interacts with their spin. We predict that the (absolute)
value of the magnetic energy of the spin $\left\vert \mathbf{\mu\cdot
B}\right\vert $ is strongly reduced. The new precession axis for the spin is
almost parallel to the axis of the nanotube and precesses with the angular
velocity of the electron. The physics of the ferromagnet is considerably modified.

\end{abstract}

Nanotubes and nanowires of both metals and semiconductors have been
extensively studied for electric charge transport. However, the electron spin
has been often ignored. How to control and manipulate the spin degree of
freedom in nanostructures is of vital importance not only for fundamental
science, but also for technological applications in micromagnetism and
spintronics. This has stimulated much research effort in the synthesis and
characterization of ferromagnetic nanowires and nanotubes. These
quasi-one-dimensional magnetic nanostructures have exhibited unique and
intriguing physical properties. As an example a number of magnetic nanotubes
of different materials show the remarkable property that their magnetic
polarization is circumferential around the axis of the tube \cite{t002},
\cite{L58} , \cite{t003}, \cite{t004}, \cite{t005}, \cite{t006} (see Fig.1).%

\begin{align*}
&
{\includegraphics[
height=2.8053in,
width=2.6368in
]%
{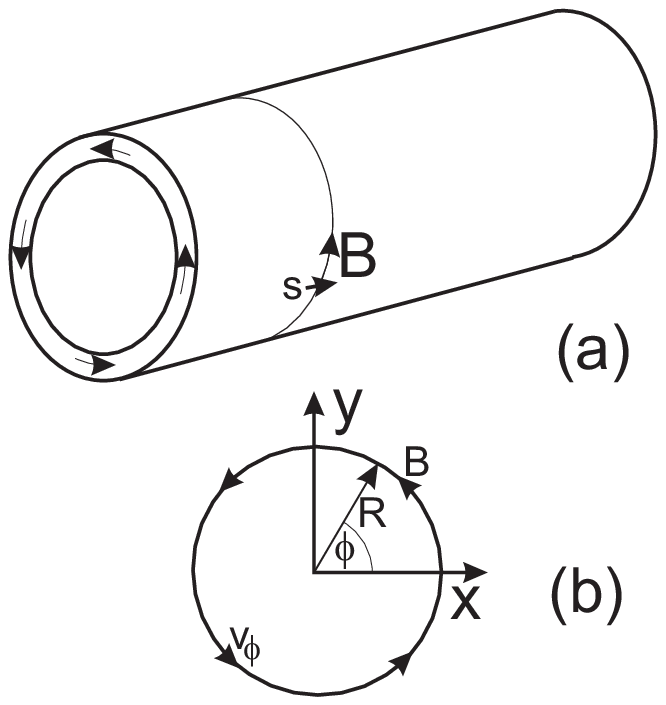}%
}%
\\
&
\begin{tabular}
[c]{l}%
Fig.1: a) A magnetic Co-nanotube with the\\
magnetization circular about the axis of the tube.\\
b) The cross section in the limit of zero thickness.
\end{tabular}
\end{align*}

In such a circular magnetic nanotube (CMNTB) the conduction electrons
experience a change of direction of the internal exchange field $\mathbf{B}$
during their propagation. This yields some interesting effects on the spin of
the electrons. For a theoretical discussion we treat the CMNTB as a tube with
zero thickness and large (infinite) mean free path of the electrons. We
consider an electron with the velocity $\mathbf{v}=v_{z}\widehat{\mathbf{z}%
}+v_{\phi}\widehat{\mathbf{\phi}}$ where $v_{z}$ is the component of the Fermi
velocity $v_{F}$ parallel to the axis and $v_{\phi}$ is the circular velocity.
The z-component of the electron velocity $v_{z}$ has no bearing on the results
of the following consideration. Therefore we set for simplicity $v_{z}=0$ and
treat the electron propagation as circular. The radius of the tube is $R$.
Then the electron circles the CMNTB with the angular frequency $\mathbf{\omega
}_{e}=\left(  v_{\phi}/R\right)  \widehat{\mathbf{z}}$.

The electron has a spin $\mathbf{s}$ and a magnetic moment of $\mathbf{\mu}$
where $\mathbf{\mu}=\gamma\mathbf{s}$ with $\gamma=-2\mu_{B}/\hbar=-e/m$ (We
set the Lande factor $g$ for the conduction electrons equal to $2$). The
circular magnetization acts as a magnetic field of strength $B_{0}$ on the
magnetic moment of the electron. In the inertial lab frame $S_{0}$ the
magnetic field causes a torque $\mathbf{\tau}$ on the magnetic moment of the
electron%
\begin{equation}
\mathbf{\tau=\mu\times B=}\gamma\mathbf{s\times B} \label{tau}%
\end{equation}
Along the circular path of the electron with the angular velocity $\omega_{e}$
the direction of the magnetic field changes. At the position $\left(
R,\phi,z\right)  $ (in cylinder coordiantes) the magnetic field is given by
\begin{equation}
\mathbf{B}=B_{0}\left(  -\sin\phi,\cos\phi,0\right)
\end{equation}
As a consequence the torque constantly changes its direction for an electron
whose position is given by $\left(  R,\omega_{e}t,z\right)  $ with
$\phi=\omega_{e}t.$ The fast angular velocity does not give the electron
enough time to precess about the direction of the local magnetic field.

In the following we treat the motion of the electron in the frame $S$ that
rotates with the electron, i.e. with the frequency $\omega_{e}$. We assign
coordinate axes $\left(  \widehat{\mathbf{x}},\widehat{\mathbf{y}}%
,\widehat{\mathbf{z}}\right)  $ to the electron position which, at $t=0$, are
equal to $\left(  \widehat{\mathbf{e}}_{1},\widehat{\mathbf{e}}_{2}%
,\widehat{\mathbf{e}}_{3}\right)  $ of the lab system. We attach these axes
$\left(  \widehat{\mathbf{x}},\widehat{\mathbf{y}},\widehat{\mathbf{z}%
}\right)  $ rigidly to the cylindrical surface of the tube (at the position of
the electron). In the next step the cylinder, electron and local axes $\left(
\widehat{\mathbf{x}},\widehat{\mathbf{y}},\widehat{\mathbf{z}}\right)  $
rotate together with frequency $\omega_{e}$ (so that $\left(  \widehat
{\mathbf{x}},\widehat{\mathbf{y}},\widehat{\mathbf{z}}\right)  =\left(
\widehat{\mathbf{r}},\widehat{\mathbf{\phi}},\widehat{\mathbf{z}}\right)  )$.

In the rotating system $S$ any vector $\mathbf{Q}$ that is constant in the
inertial system $S_{0}$ changes its direction according to
\begin{equation}
\left(  \frac{d\mathbf{Q}}{dt}\right)  _{S}=-\mathbf{\omega}_{e}%
\times\mathbf{Q}%
\end{equation}
(An analogous consideration yields the Corolis and centrifugal forces on the
surface of the earth).

For the spin this means that $d\mathbf{s}/dt$ in the rotating system is given
by%
\begin{equation}
\left(  \frac{d\mathbf{s}}{dt}\right)  _{S}=\left(  \frac{d\mathbf{s}}%
{dt}\right)  _{S_{0}}-\mathbf{\omega}_{e}\times\mathbf{s}%
\end{equation}
Here $\left(  d\mathbf{s}/dt\right)  _{S_{0}}$ is the change of the spin due
to the torque in the inertial system $S_{0}$, i.e.%
\begin{equation}
\left(  \frac{d\mathbf{s}}{dt}\right)  _{S_{0}}=\mathbf{\tau}=\gamma
\mathbf{s\times B}%
\end{equation}
In the rotating system the magnetic field is constant $\mathbf{B=}\left(
0,B_{0},0\right)  ,$ and we obtain the for $d\mathbf{s}/dt$%
\begin{equation}
\left(  \frac{d\mathbf{s}}{dt}\right)  _{S}=\gamma\mathbf{s\times
B+s\times\omega}_{e}=\gamma\mathbf{s}\times\left(  \mathbf{B+}\frac
{\mathbf{\omega}_{e}}{\gamma}\right)
\end{equation}
yielding%
\begin{equation}
\left(  \frac{d\mathbf{s}}{dt}\right)  _{S}=\gamma\mathbf{s\times B}_{eff}
\label{ds/dt}%
\end{equation}
with
\begin{equation}
\mathbf{B}_{eff}=\mathbf{B+}\frac{1}{\gamma}\mathbf{\omega}_{e}=\left(
0,B_{0},\omega_{e}/\gamma\right)  \label{Beff}%
\end{equation}
The solutions to equ. (\ref{ds/dt}) (in the system $S$) are those for a free
electron spin in a constant field $\mathbf{B}_{eff}$. The spin has a stable
constant solution (when $\mathbf{\mu}$ is parallel to $\mathbf{B}_{eff})$ and
a meta-stable solution (when $\mathbf{\mu}$ is anti-parallel to $\mathbf{B}%
_{eff})$. For a finite angle between $\mathbf{\mu}$ and $\mathbf{B}_{eff}$ the
spin performs a precession about the direction of $\mathbf{B}_{eff}$.

In components this yields%
\begin{equation}
\left(  \frac{d}{dt}\right)  _{S}\left(
\begin{array}
[c]{c}%
s_{x}\\
s_{y}\\
s_{z}%
\end{array}
\right)  =\left(
\begin{array}
[c]{c}%
\omega_{e}s_{y}-\gamma B_{0}s_{z}\\
-\omega_{e}s_{x}\\
\gamma B_{0}s_{x}%
\end{array}
\right)
\end{equation}

For the stationary solutions in the rotating system we set $\left(
d\mathbf{s}/dt\right)  _{S}=0$. This yields%
\begin{align}
s_{y}  &  =\frac{\gamma B_{0}}{\omega_{e}}s_{z}\\
s_{x}  &  =0
\end{align}
So the unit vector of the spin is in $S$%
\begin{equation}
\widehat{\mathbf{s}}=\pm\frac{1}{X}\left(  0,-\omega_{B},\omega_{e}\right)
\end{equation}
with $\hbar\omega_{B}=2\mu_{B}B_{0}$ and $X=\sqrt{\omega_{e}^{2}+\omega
_{B}^{2}}.$ The magnetic moment $\mathbf{\mu}$ is parallel to $\mathbf{B}%
_{eff}$ for the stable solution or anti-parallel in the meta-stable solution.

If the spin is not parallel or anti-parallel to $\mathbf{B}_{eff}$ then it
precesses about the effective field $\mathbf{B}_{eff}$ with a precession
frequency of
\begin{equation}
\omega_{pcs}=\gamma\left\vert \mathbf{B}_{eff}\right\vert =\sqrt{\omega
_{e}^{2}+\left(  \gamma B_{0}\right)  ^{2}}=\sqrt{\omega_{e}^{2}+\omega
_{B}^{2}}%
\end{equation}

In the rotating system the effective field $\mathbf{B}_{eff}$ is fixed with
the coordinates given by equ. (\ref{Beff}). In the inertial system
$\mathbf{B}_{eff}$ rotates with $\omega_{e}$ about the z-axis. This rotation
and the precession about $\mathbf{B}_{eff}$ have opposite senses.

\begin{align}
&
{\includegraphics[
height=2.7513in,
width=3.599in
]%
{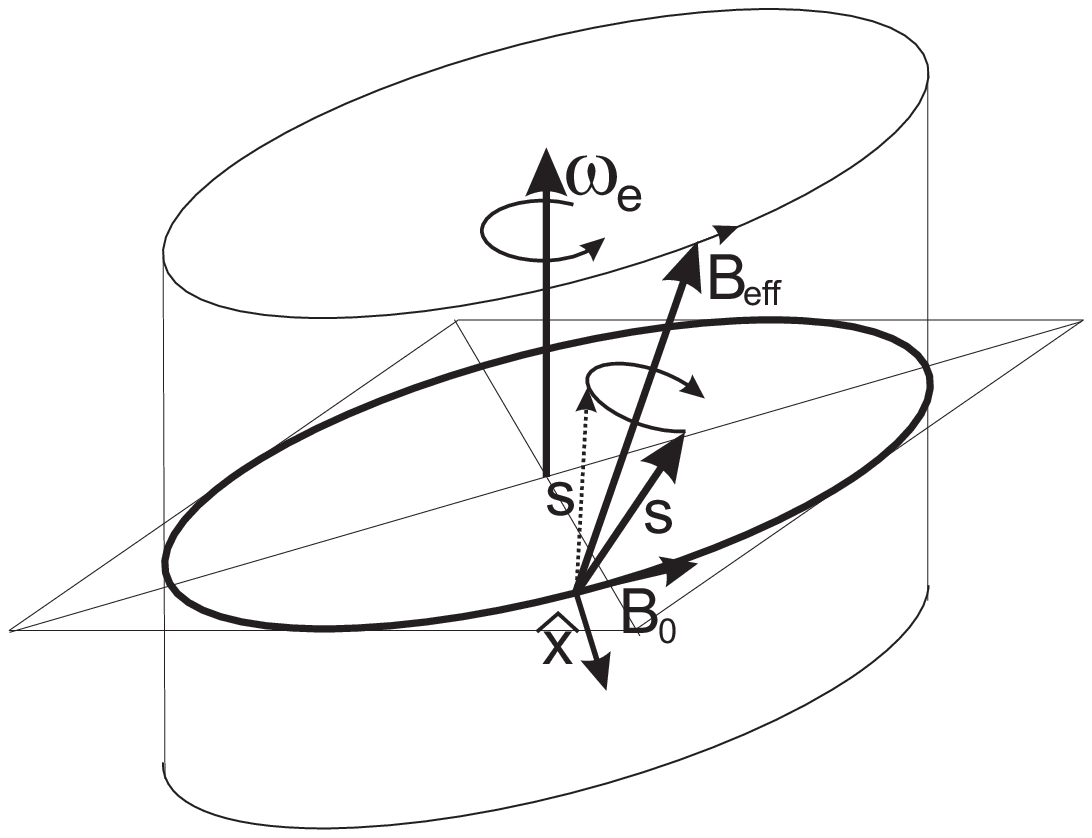}%
}%
\\
&
\begin{tabular}
[c]{l}%
Fig.2: The classical spin $\mathbf{s}$ in the frame $S$ that rotates with
$\mathbf{\omega}_{e}$.\\
The vector $\mathbf{B}_{eff}$ is fixed in the frame $S$ and the spin
$\mathbf{s}$ precesses\\
about $\mathbf{B}_{eff}$ with angular velocity $\sqrt{\omega_{e}^{2}%
+\omega_{B}^{2}}$. In the\\
inertial lab frame $S_{0}$ the precession axis $B_{eff}$ rotates\\
itself with $\mathbf{\omega}_{e}$ about the z-axis.
\end{tabular}
\end{align}

This is drawn in Fig.2. Here $\mathbf{B}_{eff}$ is the constant effective
field in the rotating frame $S.$ For $\mathbf{s}$ parallel or anti-parallel to
$\mathbf{B}_{eff}$ the spin is stationary in a stable or metastable
orientation (in $S$). If $\mathbf{s}$ has an arbitrary angle with
$\mathbf{B}_{eff}$ then $\mathbf{s}$ precesses about the axis $\mathbf{B}%
_{eff}$ with the precession frequency $\sqrt{\omega_{e}^{2}+\omega_{B}^{2}}$.
(To simplify the drawing we treated $\mathbf{s}$ and $\mathbf{\mu}$ as parallel.)

The physical picture in the \textbf{inertial lab system} $S_{0}$ is the
following. The axis of precession $\mathbf{B}_{eff}$ rotates itself with the
frequency $\mathbf{\omega}_{e}$ about the z-axis. Furthermore the spin
$\mathbf{s}$ precesses about $\mathbf{B}_{eff}$ (see Fig.2). The rotation of
the $\mathbf{B}_{eff}$ axis with frequency $\omega_{e}$ and the precession
about this axis with frequency $\sqrt{\omega_{e}^{2}+\omega_{B}^{2}}$ have
opposite senses.

For $\omega_{e}>>\omega_{B}$ the actual precession in the lab system is
approximately the difference
\begin{equation}
\omega_{pcn}\thickapprox\sqrt{\omega_{e}^{2}+\omega_{B}^{2}}-\omega
_{e}\thickapprox\frac{\omega_{B}^{2}}{2\omega_{e}} \label{om_pcn}%
\end{equation}
This is a much smaller precession frequency than $\omega_{B}=\left\vert
\gamma\right\vert B_{0}=2\mu_{B}B_{0}/\hbar$ which one would observe for a
constant magnetic field of $B_{0}\widehat{\mathbf{z}}$ in z-diection.

In addition the lowest magnetic energy of the electron magnetic moment in the
field $\mathbf{B}$ is reduced to
\begin{equation}
E_{mag}=-\mathbf{\mu\cdot B}=-\mu_{B}B_{0}\cos\theta
\end{equation}
where $\theta$ is the angle between $\mathbf{B}_{eff}$ and $\mathbf{B}$ with
\begin{equation}
\tan\theta=\frac{\omega_{e}}{\omega_{B}}%
\begin{tabular}
[c]{l}%
, \
\end{tabular}
\ \cos\theta=\frac{\omega_{B}}{\sqrt{\omega_{e}^{2}+\omega_{B}^{2}}}%
\end{equation}

A quantum theoretical treatment of this effect is desireable. The simplest
approach in the inertial lab system would be to find the right Hamiltonian.
The potential energy is straight forward $U=-\mathbf{\mu\cdot B}\left(
\phi\right)  $. There is no kinetic energy in the spin precession. But using
$U\left(  \phi\right)  $ as the Hamiltonian does not include the dynamics of
the orbiting electron. The $\phi$-dependent magnetic exchange field
$\mathbf{B}$ complicates the calculation. We apply two different approaches in
the rotating system, (i) with the local axes $\left(  \widehat{\mathbf{x}%
},\widehat{\mathbf{y}},\widehat{\mathbf{z}}\right)  $ fixed parallel to the
axis of the inertial frame and (ii) the local axes fixed to the rotating
system. The preliminary results confirm the conclusions of the classical
approach. But naturally one has to include the quantization of $\omega_{e}$
and of $\omega_{pcs}$.

\textbf{Thomas precession}: In our non-relativistic calculation we had two
coordinate systems, $\left(  \widehat{\mathbf{x}},\widehat{\mathbf{y}%
},\widehat{\mathbf{z}}\right)  $ for the rotating system $S$ and $\left(
\widehat{\mathbf{e}}_{1},\widehat{\mathbf{e}}_{2},\widehat{\mathbf{e}}%
_{3}\right)  $ for the inertial system $S_{0}$. Classically the system $S$ is
rotating with $\omega_{e}$ with respect to system $S_{0}$. However, the
rotating system $S$ experiences a time-dependent acceleration $\mathbf{a}$. In
a relativistic calculation this acceleration yields an additional precession
of the axes of the orbiting system which was first calculated by Thomas
\cite{T033} and is given by%
\begin{equation}
\mathbf{\omega}_{Th}=-\frac{1}{2c^{2}}\mathbf{v\times a}%
\end{equation}
In our case we have
\[
\mathbf{a}\mathbf{=}\left(  -R\omega_{e}^{2}\cos\left(  \omega_{e}t\right)
,-R\omega_{e}^{2}\sin\left(  \omega_{e}t\right)  ,0\right)
\]
which yields a Thomas precession about the axis of the nanotube of%
\begin{equation}
\omega_{Th}=-\frac{R^{2}\omega_{e}^{3}}{2c^{2}}%
\end{equation}
The Thomas precession changes the observed precession of the electron spin in
the inertial lab system by $\omega_{Th}$. Below we estimate the contribution
of the different terms and conclude that the Thomas precession can be neglected.

For a quantitative discussion essentially two parameters are required, the
(maximal) angular frequency $\omega_{e}$, given by the radius and the Fermi
velocity and the magnetic field acting on the conduction electrons. Here one
has two extremes cases: \newline1) Ferromagnets which are described by the
Stoner model. Here one assumes essentially only one band which is generally
the d-band of transition metal ferromagnets. The Stoner model connects the
Stoner field $B_{0}$ with the Curie temperature $T_{C}$ through the relation
$B_{0}\thickapprox k_{B}T_{C}/\mu_{B}.$ This yields fields in the range of a
few $10^{3}T$. Due to the flat d-bands the Fermi velocity is generally a
factor of 10 smaller than in (s,p)-metals. \newline2) A two-band ferromagnet
where the magnetic properties are defined by the d-electrons and the
conduction electrons are (s,p)-electrons. In this case it is more difficult to
estimate the $B_{0}$ field and the Fermi velocity. In the literature values
for the Fermi energy of spin-up and down conduction electrons are given. In
ref. \cite{T15} tunnel experiments into CoFe and NiFe alloys are evaluated
with values for the Fermi energy of $\varepsilon_{F\uparrow}\thickapprox2.2eV$
and $\varepsilon_{F\downarrow}=0.5eV$, yielding in a free electron model
$v_{F\uparrow}\thickapprox0.88\times10^{6}m/s$ and $v_{F\downarrow
}\thickapprox0.42\times10^{6}m/s$. One of the authors \cite{_018} investigated
the normal and anomalous Hall effect of amorphous Co films and obtained $0.5$
conduction electrons per Co atom in the high field region where the anomalous
Hall effect is saturated.

We use for the following estimate the value $v_{F}\thickapprox10^{6}m/s$. This
yields for a Co nanotube with the radius $R=25nm$ the (maximal) value
$\omega_{e}\thickapprox\allowbreak4.0\times10^{13}s^{-1}$. Only when the angle
between $\mathbf{B}_{eff}$ and the z-axis is small can one use the simple
relation (\ref{om_pcn}) for the precession frequency in the lab frame. In
Table I this angle $\alpha=\measuredangle\left(  \mathbf{B}_{eff}%
,\widehat{\mathbf{z}}\right)  $ is given in degrees for different values of
$B_{0}$ . Only for $B_{0}=10T$ is this angle small, and one obtains a smooth
precession of $\omega_{pcn}\approx3.9\times10^{10}s^{-1}$. For the larger
values the superposition of the rotation and precession in the rotating system
$S$ yields a complicated wobbling motion in the lab system $S_{0}$.
\begin{align*}
&
\begin{tabular}
[c]{|l|l|l|}\hline
$B_{0}$ & $\omega_{B}/\omega_{e}$ & $\alpha=\tan^{-1}\frac{\omega_{B}}%
{\omega_{e}}$\\\hline
$10T$ & $4.4\times10^{-2}$ & $2.\,\allowbreak5^{o}$\\\hline
$\ 10^{2}T$ & $4.4\times10^{-1}$ & $24^{o}$\\\hline
$10^{3}T$ & $4.4$ & $77^{o}$\\\hline
\end{tabular}
\\
&
\begin{tabular}
[c]{l}%
Table I: The ratio $\omega_{B}/\omega_{e}$ and the resulting\\
angle $\alpha$ betwenn $B_{eff}$ and the z-axis are\\
calculated for different values of $B_{0}$%
\end{tabular}
\end{align*}

For the Thomas precession frequency we obtain a value of $\omega
_{Th}\thickapprox-2.2\times10^{8}s^{-1}$. \ This value is much smaller than
precession frequency $\omega_{prc}$ and can be neglected.

To summarize our conclusion: The odd alignment of the electron spins in a
magnetic nanotube with circular magnetization has a number of interesting
effects which modify the magnetic properties. A few shall be considered here
qualitatively. In the following we assume that the mean free path of the
conduction electrons is sufficiently long so that the conduction electrons can
circle the nanotube several times before they are scattered.

(1) The ground-state energy of the circular magnetic state is increased. In a
regular ferromagnetic metal the conduction electrons align parallel or
anti-parallel to the exchange field $\mathbf{B}$ and lower their energy by
$N_{0}\left(  \mu_{B}B_{0}\right)  ^{2}$ where $N_{0}$ is the (conduction)
electron density of states per spin. In the nanotube with circular
magnetization this energy reduction is much smaller and therefore the
ground-state energy is increased by almost the same amount.

(2) This energetic effect should be particularly important for Stoner magnets.
Here the magnetic moments are band electrons (generally d-electrons) which are
not localized and possess a finite (group) velocity $\mathbf{v}_{d}\left(
\mathbf{k}\right)  =\left(  1/\hbar\right)  \partial\varepsilon_{d}\left(
\mathbf{k}\right)  /\partial\mathbf{k}$. In a magnetic field the spin-up and
-down d-electrons are shifted in opposite directions on the energy scale. The
resulting magnetization acts back on the d-moments through the Coulomb
exchange field, and the magnetization becomes Stoner-enhanced. For a
sufficiently large product of $N_{d}U$ ($N_{d}$=d-electron density of states,
$U$=Coulomb exchange energy) the d-band makes a transition into a Stoner band
magnet. This mechanism would be dramatically disturbed if the propagating
d-electrons don't align their moments in the direction of the circular
magnetization but (almost) parallel and anti-parallel to the cylinder axis. If
the $\phi$-component of $\mathbf{v}_{d}$ is sufficiently large then half the
d-electrons align their moments (roughly) parallel and the other half
anti-parallel to the cylinder axis, cancelling the exchange field. A
conclusive answer requires, of course, a detailed band structure calculation
for the Stoner system under consideration.

(3) The interaction between spin waves and the conduction electrons will be
altered. The excitation of a spin wave means the transfer of an angular
momentum $\hbar$ from a conduction electron into the spin wave. Normally this
is a simple transfer because the electron spin and the magnetization have the
same quantization direction. However, in the CMNTB the two quantization
directions are almost orthogonal to each other. The investigation of this
interaction is to be considered in the future.

(4) By covering the circular magnetic Co nanotube with another ferromagnet or
a superconductor one can investigate a cylindrical proximity effect. We expect
a considerable potential for new and interesting effects.


\begin{thebibliography}{9}                                                                                                %


\bibitem {t002}P. Landeros, O. J. Suarez, A. Cuchillo, and P. Vargas, Phys.
Rev. B79, 024404 (2009), Equilibrium states and vortex domain wall nucleation
in ferromagnetic nanotubes

\bibitem {L58}D. Li, Richard S. Thompson, G. Bergmann, J. G. Lu, Adv.
Materials 20, 4575 (2008), Template-based Synthesis and Magnetic Properties of
Cobalt Nanotube Arrays

\bibitem {t003}K. Nielsch, F. J. Casta\~{n}o, C. A. Ross, and R. Krishnan, J.
Appl. Physics 98, 034318 (2005), Magnetic properties of template-synthesized
cobalt/polymer composite nanotubes

\bibitem {t004}K. Z. Rozman, D. Pecko, L. Suhodolcan, P.l J. McGuiness, and S.
Kobe, J. Alloys Compounds 509, 551-555 (2011), Electrochemical syntheses of
soft and hard magnetic Fe50Pd50-based nanotubes and their magnetic characterization

\bibitem {t005}N. A. Usov and O. N. Serebryakova, J. Appl. Phys. 116, 133902
(2014), The peculiarities of magnetization reversal process in magnetic nanotube

\bibitem {t006}D. P. Weber, D. R\"{u}ffer, A. Buchter, F. Xue, E.
Russo-Averchi, R. Huber, P. Berberich, J. Arbiol, A. F. i Morral, D. Grundler,
and M. Poggio, NANO Letters 12, 6139-6144 (2012), Cantilever Magnetometry of
Individual Ni Nanotubes

\bibitem {T033}L. H. Thomas, Nature 117, 514 (1926), Motion of the spinning electron

\bibitem {T15}S. O. Valenzuela, D. J. Monsma, C. M. Marcus, V. Narayanamurti,
M. Tinkham, Phys. Rev. Lett. 94, 196601 (2005), Spin Polarized Tunneling at
Finite Bias

\bibitem {_018}G. Bergmann, Phys. Lett. 60A, 245 (1977), The normal Hall
effect of random close packed cobalt.
\end{thebibliography}
\end{document}